\documentclass[twocolumn,showpacs,preprintnumbers]{revtex4}
\usepackage{graphicx}
\usepackage[fleqn]{amsmath}
\usepackage{amssymb}
\usepackage{color}
\allowdisplaybreaks[2]
\newcommand{\be}{\begin{equation}}\newcommand{\ee}{\end{equation}}
\newcommand{\bea}{\begin{eqnarray}}\newcommand{\eea}{\end{eqnarray}}

\newcommand{\bvec}[1]{\mbox{\boldmath $#1$}}

\newcommand{\<}{\langle}\renewcommand{\>}{\rangle}
\newcommand{\nn}{\nonumber}
\renewcommand{\[}{\langle\!\langle}\renewcommand{\]}{\rangle\!\rangle}
\newcommand{\Ntsl}{N_{\textrm{tsl}}}
\newcommand{\Nsub}{N_{\textrm{sub}}}
\newcommand{\Niupd}{N_{\textrm{iupd}}}
\begin{document} 

\title{Nonperturbative determination of the QCD potential at 
$O(1/m)$}{DESY 06-062, RCNP-Th 06004, MKPH-T-06-12}

\author{Yoshiaki Koma$^a$, Miho Koma$^{a,b}$, Hartmut Wittig$^c$}

\affiliation{\vspace{1mm}$^a$Deutsches Elektronen-Synchrotron DESY,
Theory Group, 
D-22607 Hamburg, Germany\vspace{1mm}\\
$^b$Research Center for Nuclear Physics (RCNP), Osaka University,
Osaka 576-0047, Japan\vspace{1mm}\\
$^c$Institut f\"ur Kernphysik,
Johannes Gutenberg-Universit\"at Mainz, D-55099 
Mainz, Germany\vspace{1mm}}


\begin{abstract}
The relativistic correction to the QCD static inter-quark
potential at $O(1/m)$
is investigated nonperturbatively for the first time 
by using lattice Monte Carlo QCD simulations.
The correction is found to be comparable
with the Coulombic term of the static potential when
applied to charmonium, and amounts to 
one-fourth of the Coulombic term for bottomonium.
\end{abstract}

\maketitle

\section{Introduction}
\vspace*{-0.2cm}

Heavy quarkonia, i.e. bound states of a heavy quark and 
antiquark~\cite{Lucha:1991vn,Buchmuller:1992zf,%
Bali:2000gf,Brambilla:2004wf},
offer a unique opportunity to gain an understanding of nonperturbative QCD.
A possible way of studying such systems systematically in QCD is 
to employ nonrelativistic QCD (NRQCD)~\cite{Caswell:1985ui,Bodwin:1994jh},
which is obtained by integrating out the scale above
the heavy quark mass $m\gg \Lambda_{\rm QCD}$.
Further, by integrating out the scale $mv$, where $v$ is quark
velocity, one arrives at a framework called potential NRQCD (pNRQCD)%
~\cite{Brambilla:1999xf,Brambilla:2000gk,%
Pineda:2000sz,Brambilla:2004jw},
where the static potential emerges as the leading-order contribution,
followed by relativistic corrections in powers of $1/m$.
The potential at $O(1/m^2)$ contains the leading order spin-dependent 
corrections~\cite{Eichten:1979pu,Eichten:1980mw,Gromes:1983pm}
and the velocity-dependent 
potentials~\cite{Barchielli:1986zs,Barchielli:1988zp}.
Perturbation theory may be applied to 
the determination of these potentials to some extent.
However, since the binding energy is typically 
of the scale $mv^2$, which can be of the same order
as $\Lambda_{\rm QCD}$ 
due to the nonrelativistic nature of the system, $v\ll 1$,
as well as the fact that perturbation theory
cannot incorporate quark confinement,
it is essential to determine the potential 
nonperturbatively.
The various properties of heavy quarkonium
can be extracted by solving the Schr\"odinger equation
with these potentials.

\par
Monte Carlo simulations of lattice QCD offer a powerful tool for
the nonperturbative determination of the potentials, and 
it is the aim of this Letter to present the simulation 
result of the heavy quark potential at $O(1/m)$,
which has not been investigated so far on the lattice.
Let us denote the spatial position of the quark and antiquark as
$\vec{r}_{1}$ and $\vec{r}_{2}$ with the relative distance 
$r=|\vec{r}_{1}-\vec{r}_{2}|$ and the masses $m_{1}$ and 
$m_{2}$, respectively.
The potential is
\bea
V(r) = V^{(0)}(r) + 
\left( \frac{1}{m_{1}}+\frac{1}{m_{2} }\right )
V^{(1)}(r) +O(\frac{1}{m^2}) \; , 
\label{eqn:allpot}
\eea
where $V^{(0)}(r)$ is the static potential, usually
obtained by evaluating the expectation value of the Wilson loop.
The static potential is well parameterized by the 
Coulomb plus linear term,
\bea
V^{(0)}(r) = - \frac{c}{r} +\sigma r + \mu \;,
\label{eqn:staticpot}
\eea
where $\sigma$ is the string tension and $\mu$ a constant%
~\footnote{One may assume the L\"uscher term 
$c  = \pi/12 \approx 0.262$ at long 
distances~\cite{Luscher:1980fr,Luscher:1980ac}.}.
On the other hand, the nonperturbatively expected form of
$V^{(1)}(r)$ is not yet known, but leading-order perturbation 
theory yields  $V^{(1)}(r)= - C_{F}C_{A} \alpha_{s}^{2}/(4 r^2)$%
~\cite{Melnikov:1998pr,Hoang:1998uv,Brambilla:2000gk},
where $C_{F}=4/3$ and $C_{A}=3$ are the Casimir charges of the 
fundamental and adjoint representations, respectively
(beyond leading-order perturbation theory, 
see~\cite{Brambilla:1999xj}).

\section{Procedures}
\vspace*{-0.2cm}

\par
We work in Euclidean space in four dimensions 
on a hypercubic lattice with lattice volume 
$V=L^3 T$ and lattice spacing $a$, where 
periodic boundary conditions are imposed in all directions.
Writing the eigenstate of the pNRQCD Hamiltonian at $O(m^0)$
in the ${\bf 3} \otimes {\bf 3}^{*}$ representation of color SU(3),
which corresponds to the  static quark-antiquark state,
as $| n \> \equiv | n; \vec{r}_{1},\vec{r}_{2} \>$ 
with the energy $E_{n}(r)$ 
[e.g., $E_{0}(r)=V^{(0)}(r)$], the spectral representation of 
$V^{(1)}(r)$ is expressed as~\cite{Brambilla:2000gk,Pineda:2000sz}
\bea
V^{(1)}(r) = - \frac{1}{2}\sum_{n = 1}^{\infty}
\frac{\< 0 | g\bvec{E}(\vec{r}_{i})| n\> 
\! \cdot \!  \< n | g\bvec{E}(\vec{r}_{i})| 0\>}
{(\Delta E_{n0} )^2} \; ,
\label{eqn:v1-spectralrep}
\eea
where 
$g$ is the gauge coupling,
$\bvec{E}(\vec{r}_{i})$ denotes the electric field attached to 
the quark ($i =1$) or the antiquark ($i=2$),
and $\Delta E_{n0}\equiv E_{n}-E_{0}$  the energy gap.
It is also possible to write Eq.~\eqref{eqn:v1-spectralrep}
as the integral of the electric field strength correlator on 
the Wilson loop with respect to the relative temporal distance
between two electric fields~\cite{Brambilla:2000gk,Pineda:2000sz}.
This is, in principle, measurable on the lattice, and the result is reduced to
Eq.~\eqref{eqn:v1-spectralrep} once the spectral 
decomposition is applied by using the transfer matrix theory,
and the temporal size of the Wilson loop is taken
to infinity~\footnote{In practice, however,
the integration of the field strength correlator 
on the lattice and the extrapolation of the temporal size of 
the Wilson loop to infinity cause systematic errors, which
must be evaluated carefully.}.

\par
In our approach, the Polyakov loop correlation function (PLCF, 
a pair of Polyakov loops $P$ separated by a distance $r$) is adopted
as the quark-antiquark source instead of the Wilson loop 
for the reason discussed below.
Let us consider the field strength correlator on  the PLCF,
\bea
C(t)  \! \!&=& \! \!
\[ g^2 \bvec{E} (\vec{r}_{i},t_{1}) \! \cdot  \! \bvec{E} 
(\vec{r}_{i},t_{2}) \]_{c}\nn\\*
 \! \! & \equiv & \! \!
\[ g^2 \bvec{E} (\vec{r}_{i},t_{1}) \! \cdot  \! \bvec{E} (\vec{r}_{i},t_{2}) \]
\!- \! \[ g\bvec{E} (\vec{r}_{i})\]  \! \cdot  \! \[ g\bvec{E} 
(\vec{r}_{i})\]  \; ,\quad
\label{eqn:correlator}
\eea
where the double brackets represent the ratio of expectation value
 $\[\cdots \] =  \< \cdots \>_{PP^{*}} /  \< P P^{*}\>$, while
 $\< \cdots \>_{PP^{*}}$ implies that 
the electric field is connected to either 
of the Polyakov loop in a gauge invariant way.
The relative temporal distance of two electric field operators 
is $t=t_{2}-t_{1}$.

\par
The spectral decomposition of Eq.~\eqref{eqn:correlator} 
reads~\cite{Koma:2006a}
\bea
C(t) 
&=&
2 \sum_{n = 1}^{\infty} 
\< 0  | g\bvec{E}(\vec{r}_{i}) | n \> \! \cdot 
\! \< n | g\bvec{E}(\vec{r}_{i}) | 0 \>
e^{- (\Delta E_{n0})T/2} 
\nn\\*
&&
\times 
\cosh [(\Delta E_{n0})(T/2 -t)] 
\! + \!O(e^{-(\Delta E_{10})T})\;  ,
\label{eqn:correlator-spectralrep}
\eea
where the last term represents terms involving 
exponential factors equal to or smaller than
$\exp [-(\Delta E_{10})T]$.
Thus, once Eq.~\eqref{eqn:correlator} is evaluated via Monte Carlo simulations,
we can determine the amplitude  $|\< 0 | g\bvec{E}(\vec{r}_{i}) |n \>|^2$
and the energy gap $\Delta E_{n0}$ in Eq.~\eqref{eqn:correlator-spectralrep}
by a fit and  insert them into Eq.~\eqref{eqn:v1-spectralrep}.
It is easy to see that in the limit $T \to \infty$ we can write
Eq.~\eqref{eqn:correlator} in the integral form 
$V^{(1)}(r) = - (1/2) \lim_{\tau \to \infty} \int_{0}^{\tau} dt ~t  
C(t)$, where $\tau = \eta T$ with arbitrary $\eta  \in (0,T/2]$.

\par
The reason for using the PLCF 
is to compute Eq.~\eqref{eqn:v1-spectralrep}
with less systematic errors.
The hyperbolic cosine in Eq.~\eqref{eqn:correlator-spectralrep}
is typical for the PLCF and we can control 
the effect of the finite temporal lattice size 
on the field strength correlator automatically in the fit.
Moreover, the error term of $O(e^{-(\Delta E_{10})T})$
is already expected to be small
for a reasonable size of $T$.
By contrast, if one uses the Wilson loop at this point, 
the spectral representation is just a multi-exponential function,
and the leading error term is of $O(e^{-(\Delta E_{10}) (\Delta t)})$, 
where $\Delta t $ is the relative temporal distance between the
spatial part of the Wilson loop and the field strength operator.
Here, one cannot choose $\Delta t$ as large as $T$,
since  the temporal extent of the Wilson loop is 
limited to $T/2$
because of the periodicity of the lattice volume.

\par
The only technical problem that arises when using the PLCF 
is how to obtain a signal for the field strength correlator 
in Eq.~\eqref{eqn:correlator}, since 
the expectation value of the PLCF at zero temperature 
becomes exponentially small with increasing $r$, and
the signal is easily washed out by statistical noise.
In fact, it is almost impossible to obtain the signal of the PLCF 
at intermediate distances, say $r \approx 0.5$~fm, 
with the commonly used simulation algorithms.
However, we find that this problem can be
solved by applying the multi-level algorithm~\cite{Luscher:2001up}
with a certain modification as applied to the determination of the 
spin-dependent potentials~\cite{Koma:2005nq,Koma:2006a} 
(see also~\cite{Koma:2003gi} for a similar application).

\begin{figure}[!t]
\centering\includegraphics[width=6.8cm]{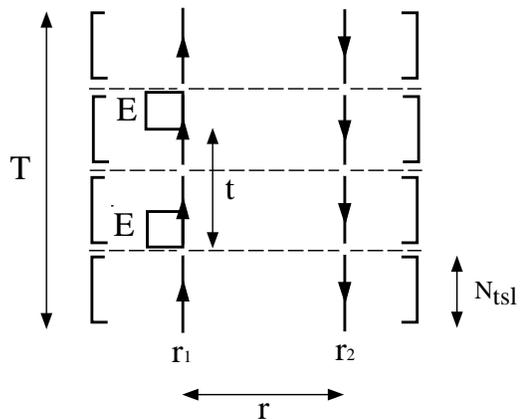}
\caption{Construction of the electric field strength correlator on 
the PLCF. Arrows at $\vec{r}_{1}$ and $\vec{r}_{2}$ represent
the Polyakov lines for the static quark and antiquark.
$[\cdots ]$ denotes the sublattice average.}
\label{fig:schematic-fsc}
\vspace*{-0.4cm}
\end{figure}

\par
The basic procedure of the multi-level 
algorithm (restricted to the lowest level) is as follows.
We first divide the lattice volume into 
several sublattices along the time direction, where
a sublattice consists of a certain number of time slices $\Ntsl$.
The number of sublattices is $\Nsub=T/\Ntsl$, 
which is assumed to be integer. In each sublattice we
take averages of the components of the correlation function
[components of the PLCF and of the field strength correlators, which
are in the ${\bf 3} \otimes {\bf 3}^*$ representation of SU(3)],
by updating the gauge field with a mixture
of heatbath~(HB) and over-relaxation~(OR) steps, 
while the spatial links on the boundary
between sublattices remain intact during the update.
We refer to this procedure as the internal update and
denote the number of internal update as $\Niupd$.
Repeating the internal update until we obtain stable 
signals for these components, we
finally multiply these averaged components to
complete the correlation function.
Thereby the correlation function is obtained for
one configuration.
For a schematic understanding, see Fig.~\ref{fig:schematic-fsc},
which illustrates the computation of the electric field strength 
correlator on the PLCF.
We then update the whole set of links without specifying any
layers to obtain another independent gauge configuration
and repeat the above sublattice averaging.
Once $\Ntsl$ and $\Niupd$ are optimized for a given gauge coupling
$\beta$ and a maximal quark-antiquark distance of interest,
the statistical fluctuations of observables 
turn out to be quite small.
Further technical details can be found in~\cite{Koma:2006a}.

\section{Results}
\vspace*{-0.2cm}

\par
Our simulations were carried out using the standard 
Wilson gauge action in SU(3) lattice gauge theory
at $\beta=6.0$ on the $20^4$ lattice 
(the lattice spacing, determined from the Sommer scale 
$r_{0}=0.5$~fm, is $a\approx 0.093$~fm~\cite{Luscher:2001up}).
One Monte Carlo update consisted of 1~HB, followed by 5~OR steps.
For practical reasons (mainly to save computer memory)
we set $\vec{r}=(r,0,0)$.
We  employed the lattice field strength operator defined by
$g a^2 F_{\mu\nu}(s) 
\equiv [U_{\mu\nu}(s)-U_{\mu\nu}^\dagger(s)]/(2i)$
at the site $s$, where $U_{\mu\nu}(s)$ are plaquette variables
and constructed the electric field by 
$g a^2 E_{i}(s)= g a^2 [F_{4i}(s)+F_{4i}(s-\hat{i})]/2$.
In order to remove self-energy contributions of the electric field we
multiplied by the conventional Huntley-Michael factor, 
$Z_{E_{i}}(r)$~\cite{Huntley:1986de}, 
which, however, only removes self-energy contributions at $O(g^2)$.
This factor, which depends on $r$ and also on the relative 
orientation of the electric field operator to $\vec{r}$,
was computed using the PLCF~\cite{Koma:2006a}. 
We obtained the value $Z_{E_{i}}(r) \approx 1.62$.
For a more precise value of $Z_{E}$,
see Ref.~\cite{Koma:2006a}.
For the chosen value of $\Ntsl =4$ we performed $\Niupd =7000$
internal updates. Our total statistics was~$N_{\rm conf}=60$.

\begin{figure}[!t]
\centering\includegraphics[width=8.8cm]{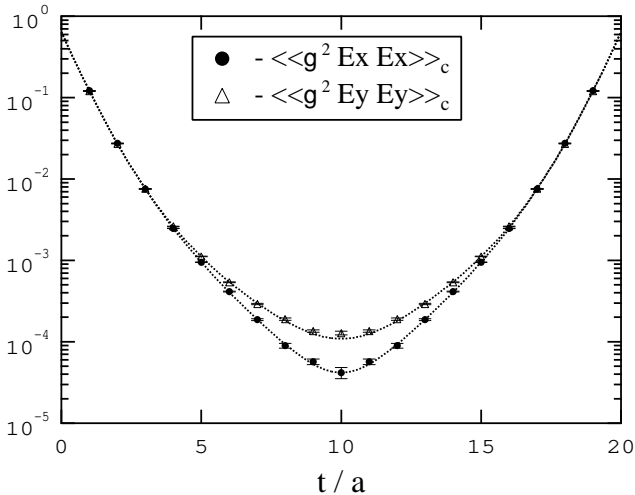}
\caption{The electric field strength correlators on the PLCF 
at $\beta=6.0$ on the $20^4$ lattice for $r/a=5$.
The dotted lines are the fit curves with $n_{\rm max}=3$ in 
Eq.~\eqref{eqn:correlator-spectralrep}.}
\label{fig:fsc}
\vspace*{-0.4cm}
\end{figure}

\par
In Fig.~\ref{fig:fsc}, we show the $C(t)$ for
the longitudinal and the transverse components,
$\[g^2 E_{x}(\vec{r}_{i},t_{1})E_{x}(\vec{r}_{i},t_{2}) \]_{c}$
and 
$\[g^2 E_{y}(\vec{r}_{i},t_{1})E_{y}(\vec{r}_{i},t_{2})\]_{c}
=\[g^2 E_{z}(\vec{r}_{i},t_{1})E_{z}(\vec{r}_{i},t_{2})\]_{c}$,
respectively, where $r/a=5$ is selected as an example.
Note that the correlators are negative.
Here, the second term of Eq.~\eqref{eqn:correlator}
can be non-zero as the electric field is even
under CP transformations.
We computed $\[g E_{i}\]$ independently and
found $\[g E_{y}\] =\[g E_{z}\] = 0$, while
$\[g E_{x}\] \ne 0$, which was then
subtracted to obtain $C(t)$.
As it is impossible to determine the amplitudes and the energy 
gaps for all~$n \geq 1$ with the limited data points,
we truncated the expansion in Eq.~\eqref{eqn:correlator-spectralrep}
at a certain $n=n_{\rm max}$.
The validity of the truncation was monitored by looking at 
$\chi^{2}$ and the stability of the
resulting potential as a function of $n_{\rm max}$, 
where $\chi^2$ was always defined with the full covariance matrix. 
We found that $n_{\rm max}=3$ was optimal
with the fit range $t/a \in [1,8]$ (equivalent to $t/a \in [12,19]$).
The systematic effect caused by the truncation 
can be checked by simulating volumes with larger values of~$T$
and by increasing $n_{\rm max}$ in the fit.
However, from the experience of evaluating 
similar field strength correlators for the spin-dependent 
potentials~\cite{Koma:2006a},
we expect that such an effect is already negligible 
compared to statistical errors, 
once three terms are included for $T=20$ at $\beta=6.0$.
Here, we employed two ways of the fit procedure; 
we fitted $\[g^2 E_{x}E_{x}\]_{c}$ and $\[g^2 E_{y}E_{y}\]_{c}$
separately, and fitted $\[ g^2 \bvec{E} \cdot \bvec{E}\]_{c}
=\[g^2 E_{x}E_{x}\]_{c}+2\[g^2 E_{y}E_{y}\]_{c}$ 
simultaneously.
The latter is based on the expectation that 
the energy gaps are the same for both correlators.
We obtained $\chi_{\rm min}^2/N_{\rm df} = 1.1$ 
for $\[ g^2 E_{x}E_{x}\]_{c}$ and $3.0$ for $\[g^2  E_{y}E_{y}\]_{c}$, 
respectively, and
the corresponding fit curves are plotted in Fig.~\ref{fig:fsc}.
$N_{\rm df}$ is the number of degrees of freedom.
The simultaneous fit yielded $\chi_{\rm min}^2/N_{\rm df} =2.2$.
In any case, the resulting potential was found to be 
the same within errors, 
which were estimated from the distribution of the 
jackknife sample of the fit parameters.
For other distances $\chi_{\rm min}^2/N_{\rm df}$ was 
smaller than in this example, and the results
of the two fit procedures
were  consistent.

\begin{figure}[!t]
\centering\includegraphics[width=8.8cm]{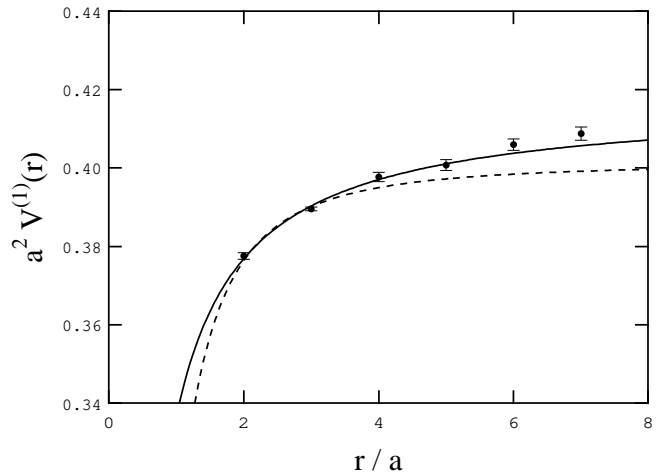}
\caption{The potential at $O(1/m)$, $V^{(1)}(r)$. 
Dashed and solid lines are the fit curves corresponding to
Eq.~\eqref{eqn:fit-pert} and Eq.~\eqref{eqn:fit-coulomb}, 
respectively.}
\label{fig:potential1}
\end{figure}

\par
We present the potential $V^{(1)}(r)$ in 
Fig.~\ref{fig:potential1}, where the result of 
the simultaneous fit is plotted.
We see an increasing behavior as a function of $r$.
We first tested whether this increasing behavior matches
the expectation from perturbation theory. 
Neglecting logarithmic corrections 
we fitted the data at $r/a \in [2,5]$ to
\bea
V_{\mbox{\footnotesize fit-1}}^{(1)}(r)
= -\frac{c'}{r^2} + \mu' \; ,
\label{eqn:fit-pert}
\eea
and found $c'=0.099(5)$ and $a^2\mu' = 0.401(1)$
with $\chi^2_{\rm min}/N_{\rm df}=6.6$,
where the fit curve is plotted in Fig.~\ref{fig:potential1}
(dashed line).
Note that if we include the data at 
$r/a=6$, $\chi^2$ becomes twice as large, 
while the fit parameters are little affected.
In order to check if this is a remnant of the perturbative behavior, 
we need data at smaller distances and perform a scaling test.
At the moment, what we can say is that the data at $r/a \gtrsim 5$
are inconsistent with a pure $1/r^2$ behavior.

\par
In trying to establish empirically the functional form of the $r$
dependence, we employed several alternative fit functions,
and among them, 
we found that
\bea
V_{\mbox{\footnotesize fit-2}}^{(1)}(r) = -\frac{c''}{r} + \mu'' \;,
\label{eqn:fit-coulomb}
\eea
can describe the behavior of  $V^{(1)}(r)$ reasonably well,
where the coefficient $c''$ has a dimension of mass. 
We took into account the data at $r/a \in [2,6]$ and
obtained $a c''=0.081(4)$ and $a^2 \mu'' = 0.417(1)$ with 
$\chi_{\rm min}^2/N_{\rm df} = 2.3$,
where the fit curve is plotted in Fig.~\ref{fig:potential1}
(solid line).

\par
As the potential $V^{(1)}(r)$ requires no matching 
coefficient~\cite{Luke:1992cs,Manohar:1997qy},
in contrast to the spin-dependent potentials at $O(1/m^2)$,
we can directly insert $V^{(1)}(r)$ into Eq.~\eqref{eqn:allpot}
and compare its relative magnitude with the static
potential $V^{(0)}(r)$
for given quark and antiquark masses.
For this purpose we may use the fit result of 
Eq.~\eqref{eqn:fit-coulomb}.
By dividing $V_{\mbox{\footnotesize fit-2}}^{(1)}(r)$ 
by the quark mass, where we set 
$m_{1}=m_{2}=m$ for simplicity,
we have a  $1/r$ term with a dimensionless coefficient $2c''/m$.
For  charmonium, $m_{c} = 1.3$~GeV,
we then find $2c''/m_{c} =  0.26(1)$, which 
is 93(5)~\% of the Coulombic coefficient of the static 
potential, $c = 0.281(5)$, in
Eq.~\eqref{eqn:staticpot}~\cite{Koma:2006a}.
For bottomonium, $m_{b} = 4.7$~GeV, 
we find $2c''/m_{b} =0.073(4)$, which is still 26(2)~\% of $c$.
It is certainly interesting to investigate the effect 
on heavy quarkonium spectroscopy.

\section{Summary}
\vspace*{-0.2cm}

\par
We have investigated the relativistic correction to the static
potential at $O(1/m)$ nonperturbatively
by using lattice QCD Monte Carlo simulations for the first time.
The key strategy here is to employ the multi-level algorithm
for measuring the field strength correlator on the PLCF and
to extract the potential by exploiting the
spectral representation of  the field strength correlators.
This method allows us to obtain the potential 
with less statistical and systematic errors.
The correction is found to be comparable
to the Coulombic term of the static potential when
applied to charmonium and to be one-fourth
of the Coulombic term for bottomonium.

\par
Finally, we note that the field strength correlator obtained here can
be used to compute one of the velocity-dependent potentials at 
$O(1/m^2)$, $V_{d}(r)$, in the parametrization of 
Refs.~\cite{Barchielli:1986zs,Barchielli:1988zp}, since the 
spectral representation of $V_{d}(r)$ consists of 
the same amplitudes and the energy gaps.
We plan to present this result as well as the other 
velocity-dependent potentials at $O(1/m^2)$ in a 
separate publication.
The first lattice result can be found in Ref.~\cite{Bali:1997am}.

\vspace{-0.7cm}
\section*{Acknowledgments}
\vspace*{-0.2cm}
 
We thank R.~Sommer, N.~Brambilla,  A.~Vairo and 
G.S.~Bali for useful discussions.
The main calculation has been performed on the NEC SX5 
at Research Center for Nuclear Physics (RCNP), 
Osaka University, Japan.
We thank H.~Togawa and A.~Hosaka for technical support.

\end{document}